\documentclass[prd,preprint,showpacs]{revtex4}
\usepackage{amsmath}
\usepackage{latexsym}
\usepackage{amsfonts}
\usepackage{graphicx}
\newcommand{\be}{\begin{equation}}
\newcommand{\ee}{\end{equation}}
\begin{document}
\title{Non-existence of Extended Holographic Dark Energy with Hubble Horizon}
\author{Yungui Gong} \email{gongyg@cqupt.edu.cn}
\affiliation{College of Mathematics and Physics, Chongqing University of Posts and
Telecommunications, Chongqing 400065, China}
\author{Jie Liu} \email{sxtyliujie@126.com}
\affiliation{College of Mathematics and Physics, Chongqing
University of Posts and Telecommunications, Chongqing 400065,
China}
\begin{abstract}
The extended holographic dark energy model with the Hubble horizon as the infrared cutoff
avoids the problem of the circular reasoning of the holographic dark energy model. We show
that the infrared cutoff of the extended holographic dark energy model cannot be the Hubble horizon
provided that the Brans-Dicke parameter $\omega$ satisfies the experimental constraint $\omega> 10^4$,
and this is proved as a no-go theorem. The no-go theorem also applies to the case in which
the dark matter interacts with the dark energy.
\end{abstract}

\preprint{arXiv: 0807.2000}

\maketitle
\tableofcontents

\section{Introduction}

The current accelerating expansion of the universe was first discovered
 in 1998 by the observation of the type Ia supernova \cite{acc1}. The
high redshift supernova Ia observation found strong evidence of
a transition from deceleration in the
past to acceleration at present \cite{riess,essence}. Evidence of the
accelerating expansion of the universe was further provided by other
complementary astronomical observations, such as the cosmic
microwave background anisotropy and the large
scale structure of the clusters of galaxies \cite{astier,sdss6,wmap3}.
As a model independent tool, the energy conditions were correctly employed
to analyze the observational data and conclude the
existence of the accelerated expansion of the universe in \cite{gong07b,gong07c}.
To explain the cosmic acceleration, an exotic energy component
with negative pressure, dubbed dark energy, is introduced. Because the
only observable effect of dark energy is through gravitational
interaction, the nature of dark energy imposes a big challenge to
theoretical physics. One simple dark energy candidate which
is consistent with current observations is the cosmological constant.
Due to the discrepancy of many orders of magnitude between
the theoretical predication and observation for the
vacuum energy, lots of dynamical dark energy models were proposed.
For a review of dark energy models, see \cite{rev}.

The holographic dark energy (HDE) model is one of the interesting dynamical dark energy
models. The HDE model is derived from the relationship between the ultraviolet (UV) and the infrared (IR)
cutoffs proposed by  Cohen {\it et al} in \cite{cohen}.
The UV-IR relationship was also obtained by Padmanabhan, arguing that the cosmological
constant is the vacuum fluctuation of energy density \cite{pad}.
Due to the limit set by the formation of a
black hole, the UV-IR relationship gives an upper bound on the zero-point
energy density $\rho_h\le 3 L^{-2}/(8\pi G)$, which means that
the maximum entropy is of the order of $S_{BH}^{3/4}$. Here $L$ is the scale of IR cutoff.
The zero-point energy
density has the same order of magnitude as the matter energy density \cite{Hsu},
and is named the HDE density by Li \cite{Li}.
However, the original HDE model with the Hubble scale as the IR cutoff failed
to explain the accelerating expansion of the universe \cite{Hsu}. Li solved the
problem by discussing the possibilities
of the particle and event horizons as the IR cutoff, and he found that
only the event horizon identified as the IR cutoff leads to a viable dark energy model \cite{Li}.
The HDE model using the event horizon as the IR cutoff was soon found to be
consistent with the observational data in \cite{gong}.
By considering the interaction between dark energy and matter in the HDE
model with the event horizon as the IR cutoff, it was shown that the
interacting HDE model realized the phantom crossing
behavior \cite{intde}. Other discussions on the HDE
model can be found in \cite{holo,gong08,hololi,holo1,pavon05,grg,horvat,holo2,
sadjadi,zhang07}

Since string theory is believed to be the theory of quantum gravity, Einstein's theory of gravity
needs to be modified according to string theory. In the low energy effective
bosonic string, the dilaton field appears naturally. The scalar degree of freedom
arises also upon compactification of higher dimensional theory. The simplest
alternative which includes the scalar field in addition to the tensor field
to general relativity is Brans-Dicke theory. Therefore, it is interesting to
discuss the HDE model in the framework of Brans-Dicke theory. That was first done by Gong
in \cite{gong04}, and the model is called the extended holographic dark energy (EHDE) model.
The EHDE model was also discussed in \cite{kim,setare,pavon,nayak,xu}.
Recently, it was claimed that the EHDE model
with the Hubble horizon as the IR cutoff could solve the dark energy problem \cite{pavon,nayak}.
The existence of the event horizon means that the universe must experience accelerated expansion, so
the HDE and EHDE models with the event horizon as the IR cutoff face the problem of circular reasoning.
If the Hubble horizon can be used as the IR cutoff in EHDE model, then the EHDE model is more
successful and interesting. In this paper, we carefully examine the EHDE model
with the Hubble horizon as the IR cutoff
and show that the model fails to solve the dark energy problem. We discuss the EHDE model with
the Hubble horizon as the IR cutoff in section II and the interacting EHDE model in section III.

\section{EHDE Model}

The Brans-Dicke Lagrangian in the Jordan frame is given by
\begin{equation}
\label{bdlagr} {\cal L}_{BD}={\sqrt{-g}\over 16\pi}\left[\phi R-
\omega\,g^{\mu\nu} {\partial_\mu\phi\partial_\nu\phi\over
\phi}\right]-{\cal L}_m(\psi,\, g_{\mu\nu}).
\end{equation}
On the basis of the flat Friedmann-Robertson-Walker metric, we get the
evolution equations of the universe
from the action (\ref{bdlagr}):
\begin{gather}
\label{jbd1} H^2+H{\dot{\phi}\over \phi}-{\omega\over
6}\left({\dot{\phi} \over \phi}\right)^2={8\pi\over 3\phi}\rho,\\
\label{jbd2}
\ddot{\phi}+3H\dot{\phi}=\frac{8\pi}{2\omega+3}(\rho-3p),
\\
\label{jbd3} \dot{\rho}+3H(\rho+p)=0.
\end{gather}
Combining the above equations, we also get
\be
\label{bdaddot}
\frac{\ddot a}{a}=H\frac{\dot \phi}{\phi}-\frac{\omega}{3}\left(\frac{\dot\phi}{\phi}\right)^2
-\frac{8\pi}{3\phi}\frac{3\omega p+(\omega+3)\rho}{3+2\omega}.
\ee
\be
\label{bdhdot}
\dot{H}=2H\frac{\dot\phi}{\phi}-\frac{\omega}{2}\left(\frac{\dot\phi}{\phi}\right)^2-
\frac{8\pi}{\phi}\frac{\omega(\rho+p)+2\rho}{2\omega+3}.
\ee

In Brans-Dicke theory, the scalar field $\phi$ takes the role of
$1/G$, so the EHDE density with the Hubble horizon
as the IR cutoff is
\be
\label{bddark}
\rho_h=\frac{3 c^2 \phi\,H^2}{8\pi}.
\ee

Let us consider the special power law solution $\phi/\phi_0=(a/a_0)^n$,
or $\dot\phi/\phi=n\,H$ first. Using equations (\ref{jbd1}), (\ref{jbd2}) and
(\ref{bdhdot}), we get the consistency condition for this solution
\begin{equation}
\label{bdconsist}
\frac{p}{\rho}=\frac{n\omega+n-1}{n\omega-3}.
\end{equation}
Substituting the power law solution into the Friedmann equation
(\ref{jbd1}), we get
\begin{equation}
\label{rhohrel}
\rho=\rho_m+\rho_h=\frac{3\phi}{8\pi}\left(1+n-\frac{\omega}{6}n^2\right)H^2.
\end{equation}
Substituting equation (\ref{bddark}) into equation (\ref{rhohrel}), we obtain
\begin{equation}
\label{rhomhrel}
\rho_m=\frac{3\phi}{8\pi}H^2\left(1+n-c^2-\frac{\omega}{6}n^2\right),
\end{equation}
and
\begin{equation}
\label{wmwhratio}
r=\frac{\Omega_m}{\Omega_h}=\frac{1+n-c^2-\omega n^2/6}{c^2}.
\end{equation}
Therefore, this solution is the tracking solution in which the dark energy tracks the matter.
When $n=3/\omega$, the ratio $r$ reaches the maximum value
\be
\label{omolmax1}
r_{max}=\frac{1-c^2+3/(2\omega)}{c^2}.
\ee
On the other hand, if
\begin{equation}
\label{nomega}
n = \frac{3\pm \sqrt{9+6\omega(1-c^2)}}{\omega},
\end{equation}
then we get the dark energy dominated solution with $r=0$.
For the dark energy dominated solution, the deceleration parameter is
\be
\label{bdqpar1}
q=2+\frac{3\pm\sqrt{9+6\omega(1-c^2)}}{\omega}\pm \frac{\sqrt{3}\, c^2}{\sqrt{3+2\omega(1-c^2)}}.
\ee
If $c^2=1$, we get $n=0$ and $n=6/\omega$. So $q=1$ and $q=3+6/\omega$, respectively.
To get late time acceleration,
we require $-2<\omega<0$ \cite{gong04} which is in violation of the
current experimental constraint $\omega>10^4$ \cite{will,bertotti}.
If $c^2\neq 1$, then $q<0$ requires that $c^2$ is very close to $1+3/(2\omega)$
when $\omega\gg 1$. However, equation (\ref{wmwhratio}) tells us
that $r\sim 0$ if $c^2\sim 1+3/(2\omega)$. This means that we cannot
recover the early matter dominated epoch.
Therefore, if the Brans-Dicke scalar field takes the power law form $\phi/\phi_0=(a/a_0)^n$,
the EHDE model with the Hubble horizon as the IR cutoff exists
only when $\omega<0$.

We may wonder whether the EHDE model with the Hubble horizon as the IR cutoff
exists if we consider more general solutions. In order to analyze the system,
let us take $y=H^{-1}\dot\phi/\phi$; then equation (\ref{jbd1}) becomes
\begin{equation}
\label{rhohre2}
\rho=\rho_m+\rho_h=\frac{3\phi}{8\pi}H^2\left(1+y-\frac{\omega}{6}y^2\right).
\end{equation}
Substituting equation (\ref{bddark}) into equation (\ref{rhohre2}), we get
\begin{equation}
\label{rhomhre2}
\rho_m=\frac{3\phi}{8\pi}H^2\left(1+y-c^2-\frac{\omega}{6}y^2\right),
\end{equation}
and
\begin{equation}
\label{wmwhratio1}
r=\frac{\Omega_m}{\Omega_h}=\frac{1+y-c^2-\omega y^2/6}{c^2}.
\end{equation}
Since $\rho_m\ge 0$ and $\omega>0$, so
\be
\label{yregion}
\frac{3-\sqrt{9+6\omega(1-c^2)}}{\omega}\le y\le \frac{3+\sqrt{9+6\omega(1-c^2)}}{\omega},
\ee
and
\be
\label{c2cond2}
c^2\le 1+\frac{3}{2\omega}.
\ee
Again when $y=3/\omega$, $r$ reaches the maximum value,
\be
\label{omovmax}
r_{max}=\frac{1-c^2+3/(2\omega)}{c^2}.
\ee
To recover the matter dominated universe, either $c^2$ or $\omega$ must be very small.
Let $x=\ln a$; with the help of equations (\ref{jbd1}) and (\ref{bdhdot}), equation (\ref{jbd2})
is rewritten as
\be
\label{dydxeq}
y'=\frac{[\omega y^2-6y-6(1-c^2)][(\omega+1)y-1]}{2[3+2\omega(1-c^2)]},
\ee
where $y'=dy/dx$. Take $y'=0$; we get three fixed points
\be
\label{bdfxpts}
y_{1c}=\frac{3+\sqrt{9+6\omega(1-c^2)}}{\omega},\quad
y_{2c}=\frac{3-\sqrt{9+6\omega(1-c^2)}}{\omega},\quad
y_{3c}=\frac{1}{1+\omega}.
\ee

From the definition (\ref{bddark}) of the HDE and
the energy conservation equation (\ref{jbd3}) of the dark energy,
we get the equation of state parameter of the dark energy
\be
\label{wlambda}
w_h=\frac{3-(4\omega+3)\,y+\omega(\omega+1)\,y^2}{3[3+2\omega(1-c^2)]}.
\ee
From equation (\ref{bdaddot}), we get the deceleration parameter
\be
\label{bdqpar2}
q=-\frac{\ddot a}{a H^2}=\frac{1}{2}+\frac{\omega(\omega+1)\,y^2-2\omega(1+c^2)\,y+3}{2[3+2\omega(1-c^2)]}.
\ee
If $c^2\le \sqrt{3(1+1/\omega)}\,-1$, then $q$ is never less than $1/2$.

Substituting the fixed points $y_{1c}$ and $y_{2c}$ into equation (\ref{wmwhratio1}),
we get $r=0$.
These two fixed points correspond
to the power law solution (\ref{nomega}) discussed above.
The deceleration parameters are given by equation (\ref{bdqpar1}).

For the first fixed point $y_{1c}$, equation (\ref{bdqpar1})
tells us that $q>2$. This fixed point corresponds
to the deceleration solution. By linearizing equation (\ref{dydxeq}) around the fixed point $y_{1c}$, we get
\be
\label{liny1c}
y'=\frac{2\sqrt{9+6\omega(1-c^2)}\,[3+2\omega+(\omega+1)\sqrt{9+6\omega(1-c^2)}\,]}{\omega}y.
\ee
Therefore, the fixed point $y_{1c}$ corresponds to an unstable fixed point which is not interesting.

For the second fixed point $y_{2c}$, equation (\ref{bdqpar1}) tells us that $q$ can be negative
when $c^2\approx 1+3/(2\omega)$. However, when $c^2\approx 1+3/(2\omega)$,
$y\approx y_{1c}\approx y_{2c}$ and $r=0$. The universe is always in
the dark energy dominated era and the matter dominated era does not exist.
By linearizing equation (\ref{dydxeq})
around the fixed point $y_{2c}$, we get
\be
\label{liny2c}
y'=-\frac{2\sqrt{9+6\omega(1-c^2)}\,[3+2\omega-(\omega+1)\sqrt{9+6\omega(1-c^2)}\,]}{\omega}y.
\ee
When $(2\omega+3)(3\omega+4)/6(\omega+1)^2< c^2\le 1+3/2\omega$, the fixed point corresponds
to a stable fixed point.

\begin{figure}[htp]
\centering
\includegraphics[width=16cm]{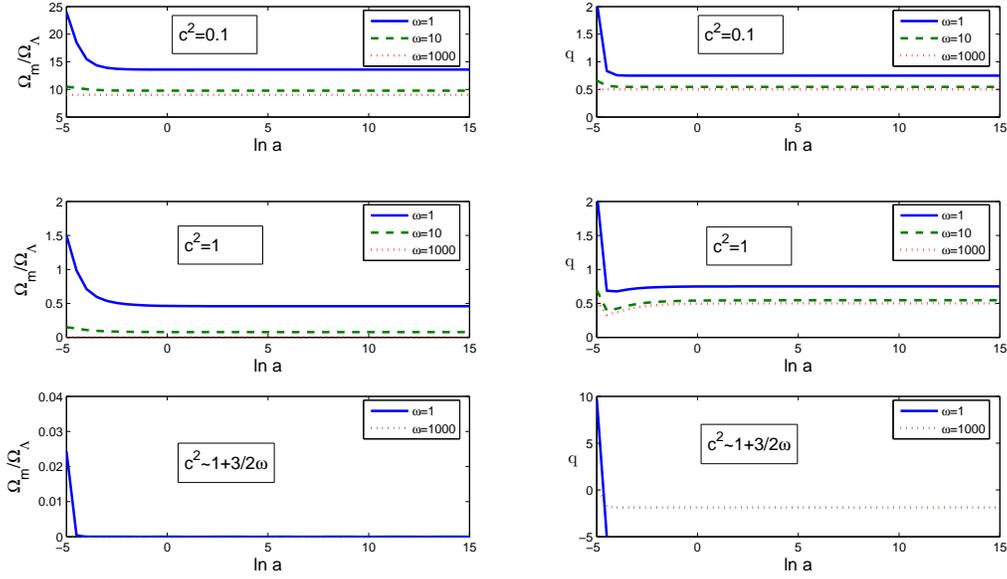}
\caption{The evolutions of $\Omega_m/\Omega_\Lambda$ and $q$ for different $c^2$ and $\omega$.}
\label{fxptehde}
\end{figure}

For the third fixed point,
we get
\be
\label{omoly3c}
r=\frac{1-c^2}{c^2}+\frac{6+5\omega}{6(1+\omega)^2 c^2}>\frac{1-c^2}{c^2},
\ee
and the deceleration parameter
\be
\label{qpary3c}
q=\frac{1}{2}+\frac{1}{2(\omega+1)}> \frac{1}{2}.
\ee
This fixed point corresponds to the deceleration solution too.
By linearizing equation (\ref{dydxeq}) around the fixed point $y_{3c}$, we get
\be
\label{liny3c}
y'=-\frac{(2\omega+3)(3\omega+4)-6(\omega+1)^2\,c^2}{2(\omega+1)[3+2\omega(1-c^2)]}y.
\ee
When $c^2 < (2\omega+3)(3\omega+4)/6(\omega+1)^2$, the fixed point becomes the stable fixed point.
The analysis of the fixed points is summarized in table \ref{fxptstab}.
To better understand the above discussion, we take the parameters
 $c^2=0.1$, $c^2=1$, $c^2\approx 1+3/2\omega$,
$\omega=1$, $\omega=10$ and $\omega=1000$, and then solve equation (\ref{dydxeq}) numerically.
The evolutions of $r=\Omega_m/\Omega_h$ and $q$ are plotted in figure \ref{fxptehde}.
The results in figure \ref{fxptehde} support the above analysis.

\begin{table}[htp]
\begin{tabular}{clll}
\hline
Points & \ \ \ Stability & \ \ \ \  $r$ & \ \ \ \ $q$ \\
\hline
$y_{1c}$ & \ \ \ Unstable & \ \ \ \ 0 & \ \ \ \  $>2$\\
$y_{2c}$ & \ \ \ Stable if $c^2>(2\omega+3)(3\omega+4)/6(\omega+1)^2$ & \ \ \ \ 0 & \ \ \ \ Can be negative \\
$y_{3c}$ & \ \ \ Stable if $c^2<(2\omega+3)(3\omega+4)/6(\omega+1)^2$ & \ \ \ \ $>(1-c^2)/c^2$ & \ \ \ \ $>1/2$ \\
\hline
\end{tabular}
\caption{The property of the fixed points for the EHDE model without interaction.}
\label{fxptstab}
\end{table}

From the above discussion, we know that the
transition from deceleration in the past which is matter dominated to acceleration
at present which is dark energy dominated does not happen in the EHDE model
with the Hubble horizon as the IR cutoff. The result is somehow expected because
Brans-Dicke theory reduces to Einstein theory when $\omega\rightarrow \infty$.
In Einstein theory, we know that the HDE model with the Hubble horizon as the IR
cutoff does not exist. So we expect that the EHDE model with the Hubble horizon as the IR
cutoff not to exist either if $\omega\gg 1$. Therefore, we
have the following no-go theorem: There is no viable EHDE model
with the Hubble horizon as the IR
cutoff if the Brans-Dicke parameter $\omega$ satisfies the
experimental constraint $\omega> 10^4$. This no-go theorem also applies
to the situation when we consider the interaction between dark
matter and dark energy.

\section{Interacting EHDE model}

By introducing the interaction between dark matter and
dark energy, the conservation equations become
\be
\label{rmcons}
\dot\rho_m+3H\rho_m=Q,
\ee
\be
\label{rxcons}
\dot\rho_h+3H(1+w_h)\rho_h=-Q,
\ee
where $Q$ stands for the interaction term. We take $Q=\Gamma H\rho_h$ with $\Gamma>0$
being the interaction rate. Because the Friedmann equation (\ref{jbd1}) is unchanged, the
ratio $r$ between $\Omega_m$ and $\Omega_h$ still satisfies equation (\ref{wmwhratio1}) and
the range of the variable $y$ is still restricted by equation (\ref{yregion}).
The above discussion tells us that for $\omega \gg 1$, we need to require $c^2\ll 1$ to get
an early matter dominated universe.
The differential equation of the variable $y$ now becomes
\be
\label{ydereq3}
y'=\frac{\omega(\omega+1)y^3-(6+7\omega)y^2+2[(3+3\omega-\omega\Gamma)c^2-3\omega]y+6(1-c^2+\Gamma c^2)}{2[3+2\omega(1-c^2)]},
\ee
The fixed points are derived by setting $y'=0$. In this case, there are no analytical expressions for the fixed points.
We need to find them numerically. Note that the fixed points
$y_{1c}$ and $y_{2c}$ in equation (\ref{bdfxpts}) are no longer
fixed points when the interaction between dark components is present.
In other words, the late time attractors do not correspond to $r=0$ due to the interaction.
The equation of state parameter $w_h$ is
\be
\label{wleq3}
w_h=\frac{\omega(\omega+1)y^2-(4\omega+3)y+3-(2\omega+3)\Gamma}{3[3+2\omega(1-c^2)]}.
\ee
The deceleration parameter is
\be
\label{qeq4}
q=\frac{1}{2}+\frac{\omega(\omega+1)\,y^2-2\omega(1+c^2)\,y+3-2\omega c^2\Gamma}{2[3+2\omega(1-c^2)]}.
\ee
With the extra term $\Gamma$, it is now easier to get $q<0$.

\begin{figure}[htp]
\centering
\includegraphics[width=14cm]{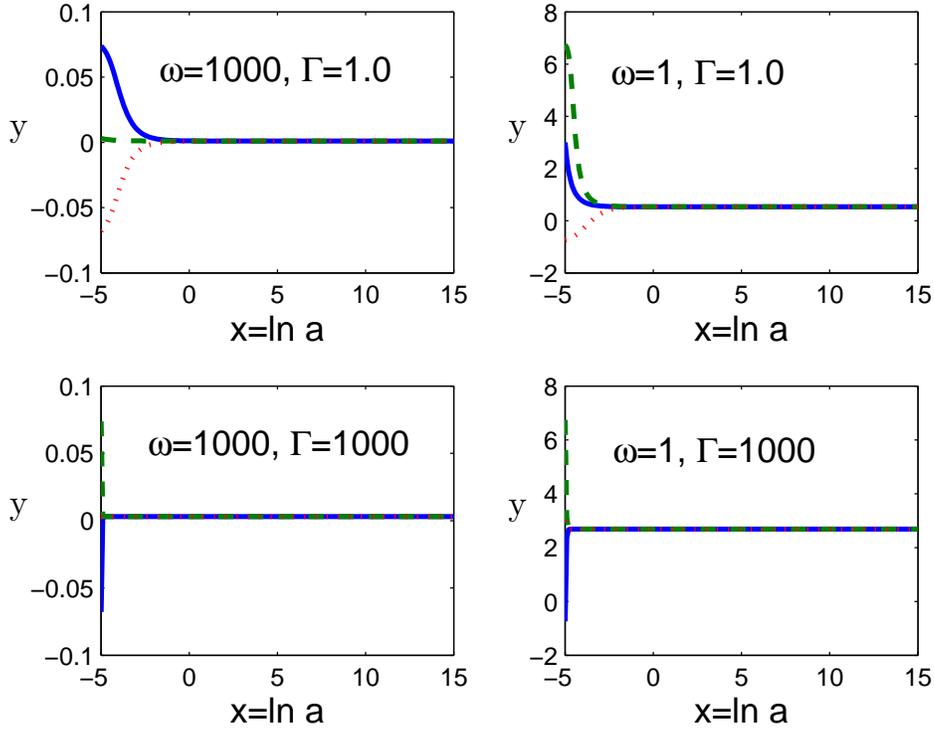}
\caption{The dynamical evolutions of $y$ for different $\Gamma$ and $\omega$, we take $c^2=0.1$.}
\label{ehdehq}
\end{figure}

\begin{figure}[htp]
\centering
\includegraphics[width=12cm]{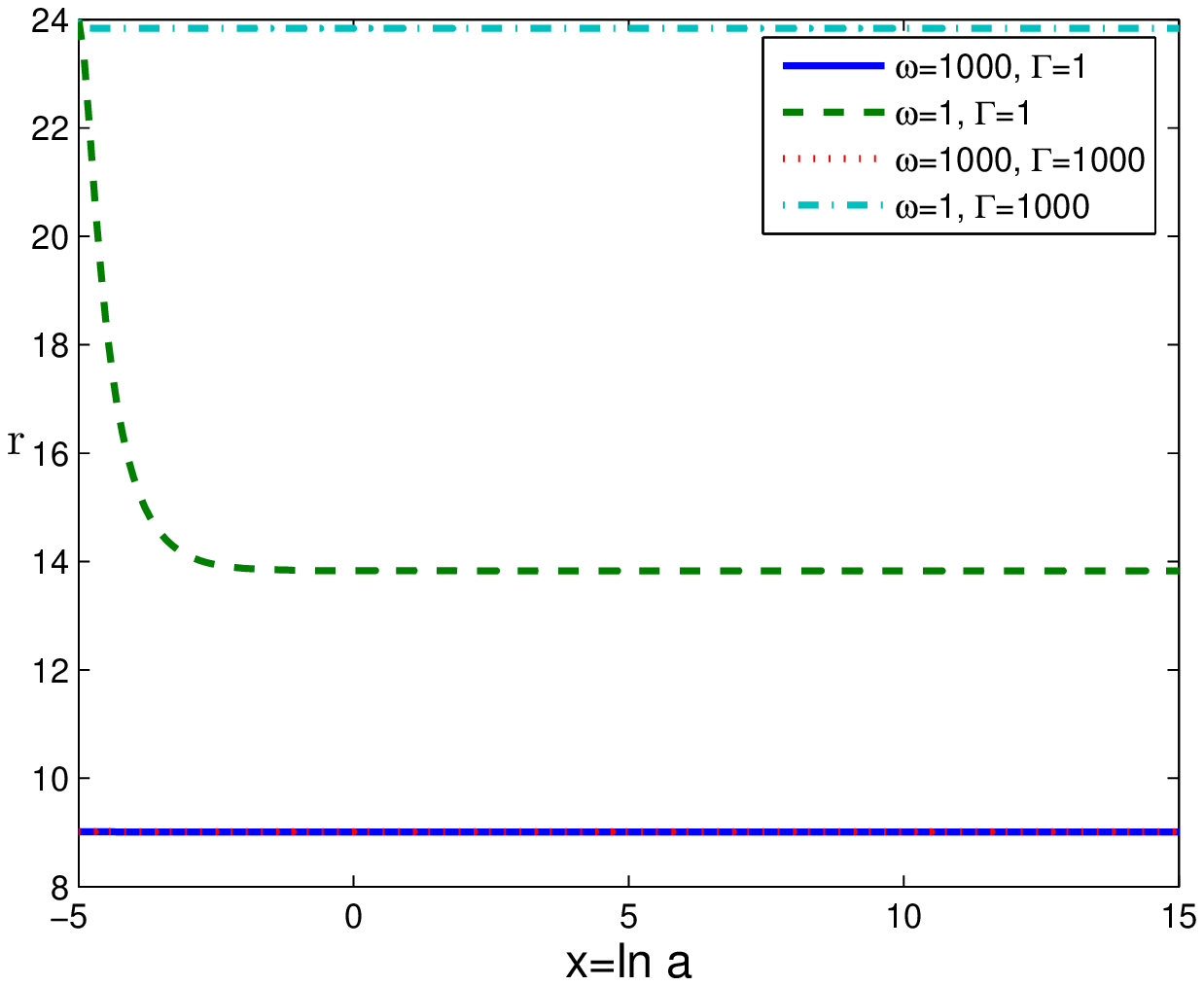}
\caption{The dynamical evolutions of $r$ for different $\Gamma$ and $\omega$, we take $c^2=0.1$.}
\label{ehdehqa}
\end{figure}

\begin{figure}[htp]
\centering
\includegraphics[width=14cm]{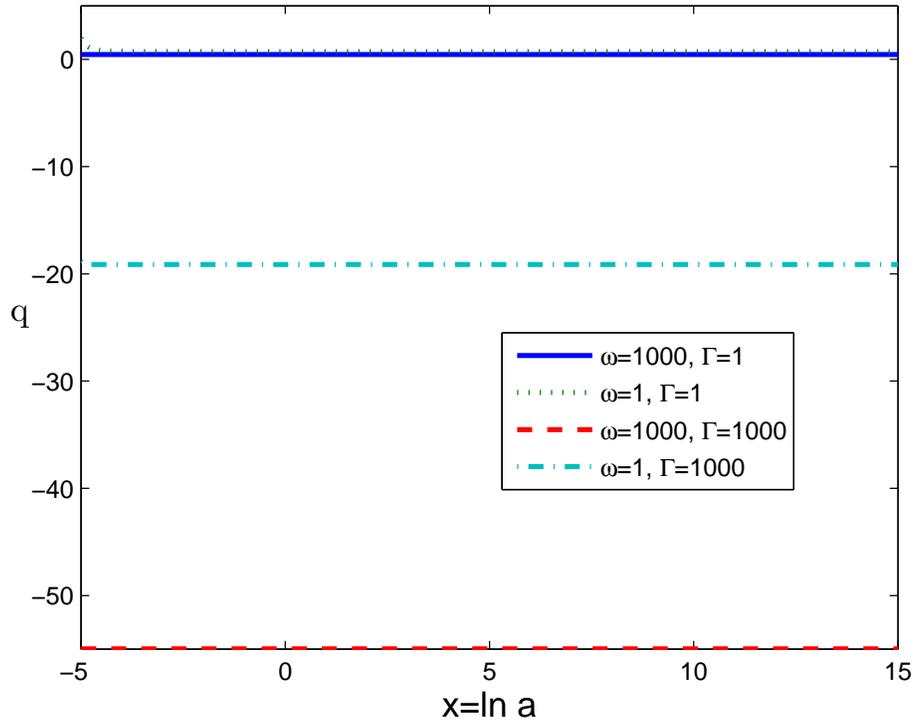}
\caption{The dynamical evolutions of $q$ for different $\Gamma$ and $\omega$, we take $c^2=0.1$.}
\label{ehdehqb}
\end{figure}

To solve equation (\ref{ydereq3}), we need to specify $\omega$, $c^2$ and $\Gamma$. From the above
discussions, it is necessary that $c^2\ll 1$ so that it is possible for the universe to have experienced
the transition from the matter dominated to the dark energy dominated case. If $\Gamma$ is order unity, then
$\Gamma c^2\ll 1$ and the effect of the interaction becomes negligible. So the late time attractor
is approximately the same as that without interaction, i.e., the fixed point is $y_c\sim 1/(\omega+1)$.
However, the dark energy would not dominate if $\Gamma\,c^2\ll 1$ as shown in equation (\ref{omoly3c}).
To illustrate the point, we take
$c^2=0.1$ and assume that $\Gamma$ is a constant for simplicity. In figure \ref{ehdehq}, we show
the dynamical behavior of $y$ for $\omega=1$ and $\omega=1000$, and $\Gamma=1$ and $\Gamma=1000$.
If $\Gamma=1$, the fixed points are $y_c=0.001$ for $\omega=1000$ and $y_c=0.53$ for $\omega=1$.
The results support our argument that $y_c\approx 1/(1+\omega)$. If $\Gamma=1000$, the fixed points are $y_c=0.00295$
for $\omega=1000$ and $y_c=2.687$ for $\omega=1$. Now the fixed points are $y_c\approx 3/\omega$.
In figures (\ref{ehdehqa}) and (\ref{ehdehqb}), we show the evolution of $r$ and $q$ for the same
choices of $\omega$ and $\Gamma$. From figure (\ref{ehdehqa}), it is clear that there is no dark
energy dominated period. We also see that $r$ is almost independent of $\Gamma$ if $\omega\gg 1$.
This can be easily understood. If $\omega\gg 1$, then
$-\sqrt{6(1-c^2)/\omega}\alt y\alt \sqrt{6(1-c^2)/\omega}$. So $y$ is almost zero. Furthermore,
the fixed point $y_c\sim 1/\omega$, so $r\sim (1-c^2)/c^2$ which is independent of $\Gamma$.
From figure (\ref{ehdehqb}), we see that there is no transition from deceleration to acceleration.

In conclusion, the no-go theorem also applies for the interaction case. Whether there is interaction
between the dark components or not, the EHDE model with the Hubble horizon as the
IR cutoff is not a viable dark energy model if the Brans-Dicke parameter $\omega$ satisfies the experimental
constraint $\omega\gg 1$. The no-go theorem breaks down if $\omega<0$. For a more general Brans-Dicke
theory with variable $\omega$, the current solar system constraint can be relaxed \cite{mota}. If $\omega$
is a function of the scalar field and $\omega(\phi)$ increases with time, we may get different result.
The HDE model in the context of a general scalar-tensor theory of gravity will be considered in future work.

\begin{acknowledgments}
The work is supported by NNSFC under grant No. 10605042. Y.G. Gong thanks the hospitality of the
Abdus Salam International Center for Theoretical Physics where part of the work was done.
\end{acknowledgments}

\end{document}